\documentclass[graybox]{svmult}

\usepackage{mathptmx}       
\usepackage{helvet}         
\usepackage{courier}        
\usepackage{type1cm}        
\usepackage{makeidx}         
\usepackage{graphicx}        
\usepackage{multicol}        
\usepackage[bottom]{footmisc}

\usepackage[normalem]{ulem} 
\usepackage[colorlinks=true,
           	 linkcolor=green,
            	 citecolor=green]{hyperref}

\newcommand{\m}{\mbox{m}}

\makeindex             

\begin{document}

\title*{Modelling sea ice and melt ponds evolution: sensitivity to microscale heat transfer mechanisms}
\titlerunning{Melt ponds evolution: sensitivity to microscale heat transfer mechanisms} 
\author{Andrea Scagliarini, Enrico Calzavarini, Daniela Mansutti and Federico Toschi}
\institute{Andrea Scagliarini \at IAC-CNR, Institute for Applied Mathematics `M. Picone', Via dei Taurini 19, 00185 Rome, Italy \email{andrea.scagliarini@cnr.it}
  \and Enrico Calzavarini \at Universit\'e de Lille, Unit\'e de M\'ecanique de Lille, UML EA 7512, F 59000 Lille, France\\ \email{enrico.calzavarini@polytech-lille.fr}
  \and Daniela Mansutti \at IAC-CNR, Institute for Applied Mathematics `M. Picone', Via dei Taurini 19, 00185 Rome, Italy  \email{d.mansutti@iac.cnr.it}
  \and Federico Toschi \at TU/e, Eindhoven University of Technology, P.O. Box 513, 5600 MB Eindhoven, The Netherlands and IAC-CNR, Institute for Applied Mathematics `M. Picone', Via dei Taurini 19, 00185 Rome, Italy \email{f.toschi@tue.nl}}
%
%
\maketitle

\abstract{We present a mathematical model describing the evolution of sea ice and meltwater during summer.
  The system is described by two coupled partial differential equations for the ice thickness $h$
  and pond depth $w$ fields. We test the sensitivity of the model to variations of 
  parameters controlling fluid-dynamic processes at the pond level, namely the variation of turbulent heat flux
  with pond depth and the lateral melting of ice enclosing a pond.
  We observe that different heat flux scalings determine different rates of total surface ablations,
  while the system is relatively robust in terms of probability distributions of pond surface areas.
  Finally, we study pond morphology in terms of fractal dimensions, showing that the role
  of lateral melting is minor, whereas there is evidence
  of an impact from the initial sea ice topography.}

\section{Introduction} \label{sec:intro}

The Arctic Ocean is characterised by the presence of ice, formed from the freezing of oceanic water.
Such layer of sea ice is a key component of the Earth Climate System \cite{Hunke2010,Notz2012}, for it represents
a sort of `boundary condition' for heat, momentum and mass exchange between ocean and atmosphere at high latitudes
\cite{CattleCrossley,Ebert1995,MaykutMcPhee,Tsamados2014} and plays a crucial role in the salinity balance in the
ocean \cite{Vancoppenolle2009a,Vancoppenolle2009b}, thus affecting also the thermohaline circulation \cite{Mauritzen}.
Moreover, sea ice turns out to be a sensitive indicator of climate change: during the last few decades its
average thickness and extent decreased significantly \cite{Kwok2009,Stroeve2012,Laxon2013}.
This decrease is two-way coupled with global warming, which
shows up particularly striking in the Arctic, via the so called ice-albedo feedback.
Sea ice, in fact, has a large albedo as compared to open oceanic waters, i.e. it reflects a high fraction
of the incident solar radiation, while water absorbes it, thus favouring warming.
The warmer the Earth surface the more ice melts, the lower gets the global albedo. 
The variability of sea ice emerges as the result of many processes acting on different time scales.
The energy budget involving incoming and outgoing radiation \cite{MaykutUntersteiner,EbertCurry,EisenmanWettlaufer},
the melting phase transition \cite{Steele1992,BitzLipscomb},
the transport of water through ice porous structure \cite{FreitagEicken,FelthamMushy,WellsMushy,TurnerHunke}, the rheology of
internal stresses \cite{FelthamRheo,HunkeDukowicz,Tsamados2013,Rabatel2018}, the transport forced by couplings with
ocean and atmosphere \cite{Steele1997,Schroeder2003,Rampal2009,Rampal2011,Petty2013,Tsamados2018},
all these make sea ice an extremely complex system and its theoretical modelling a challenge
\cite{Hunke2010,Notz2012,HunkeCryo,Massonnet}.\\
An important role in the ice-albedo feedback is played by the presence, on the ice surface,
of melt ponds \cite{FettererUntersteiner,Perovich2002aerial}: during
summer both the snow cover and the upper surface of sea ice melt and, as a consequence, meltwater may accumulate in depressions
of the ice topography (thus forming ponds).
The albedo of a melt pond ranges between $\sim 0.1$ and $\sim 0.5$ \cite{Hanesiak}, while for ice between
$\sim 0.4$ and $\sim 0.8$ \cite{FettererUntersteiner}. The average albedo
for ponded ice is, then, lower than for the unponded one \cite{Perovich2002albedo}.
The evolution of melt ponds and of their distribution over the sea ice surface is, therefore,
a key ingredient to be accounted for in realistic models of sea ice.
It has been indeed suggested that a missing or improper inclusion of melt ponds could be the cause
of overestimation, by certain general circulation models (GCMs),
of the September sea ice minumum \cite{Flocco2012,Schroeder2014}.
For climatological temporal scales, it is important to get an accurate enough knowledge of the pond depth
and surface area distributions, since these ones impact on the
radiation budget; the rate of heat transfer through the ice pack, moreover, depends on the dynamics of meltwater,
which, despite the average shallowness of ponds, can be turbulent \cite{TaylorFeltham}.\\ 
The complexity of the melt-pond-covered sea ice system resides exactly in this intrinsic multiscale nature.
Borrowing terms from Condensed Matter Physics, one can say that a modellistic approach may be
tackled, at least, at three level of description: a {\it microscopic} level, where the focus is on the
``atoms'' of the system, the single pond and the fluid dynamics inside it, as done in, e.g.,
\cite{SkyllingstadPaulson,Enrico}; a {\it mesoscopic} level, where the evolution of many ponds
is considered, coupled with the evolution of a resolved sea ice topography
\cite{Luethje,Luethje2,SkyllingstadEtAl,ScottFeltham}; and, finally,
a {\it macroscopic} level, on scales of climatological interest, where sea ice dynamics is described in terms
of an ice thickness distribution (ITD) \cite{Thorndike,CICE,LIM},
and melt ponds need to be parametrized \cite{Flocco2012,FloccoFeltham,Flocco2010}.
The aim of this contribution is twofold. We will propose a {\it mesoscopic} model
(in the sense explained above) and employ it to assess the
sensitivity of the melt-ponds-covered sea ice system to different modelling of certain dynamical processes
occurring at the single pond {\it microscopic} level.\\
The paper is organized as follows:
in section \ref{sec:ourmm} we introduce the proposed
mathematical model and its numerical implementation; in section \ref{sec:results} the main results are illustrated and discussed,
while concluding remarks and research outlooks are left to section \ref{sec:concl}.

\section{The mathematical model}\label{sec:ourmm}

The physical processes that occur within the ice pack and lead to variation of the sea ice thickness, can
be grouped essentially into two categories: thermodynamic and mechanical. Thermodynamic processes are those
related to the radiative budget; the fraction of incoming radiation that is absorbed is spent to increase
the surface temperature and to melt ice.
Mechanical deformations of sea ice are induced by ocean and wind stresses. These can drive sea ice
transport, as well as elasto-plastic deformations in the pack, giving rise to events such as ridging and rafting
\cite{Hunke2010}. Since we are interested in simulating processes involving ice melting and
  meltwater dynamics, we will neglect sea ice
transport and mechanical terms (despite they can act on time scales comparable to melting in summer). 
As ice melts, meltwater is formed and transported, by sliding over the ice topography and seepage through
its porous structure. It will eventually concentrate in {\it local minima} of the ice topography, forming melt ponds.

\subsection{The sea-ice-thickness/melt-pond-depth system} \label{subsec:equations}

We consider, therefore, the evolution of the ice (of density $\rho_i$) thickness field $h(\mathbf{x},t) \geq 0$ and the
meltwater (of density $\rho_w$) pond depth field $w(\mathbf{x},t) \geq 0$ (with $\mathbf{x} \in \Omega \subset R^2$),
whose dynamical equations read:

\begin{eqnarray}\label{eq:SM1}
\partial_t h & = & -f \\ \nonumber
\partial_t w & = & -\nabla \cdot (\mathbf{u} w) + \frac{\rho_i}{\rho_w}f  - s,
\end{eqnarray}
where $f$, $\mathbf{u}w$ and $s$ represent, respectively, the melting rate,
the meltwater flux (per unit cross-sectional area) and the seepage rate,
which are, in general, functionals of $h$ and $w$.


Similar mesoscopic models based on the evolution of $h$ and $w$ have been proposed in the past \cite{Luethje,ScottFeltham}.
Here, the original contributions to the modelling are in the parametrization of
fluid-dynamics processes, in particular the water transport term and, more importantly,
the vertical and lateral melt-rate term in turbulent flow conditions, which we will describe in detail in the following.

\subsubsection{Melting rate}

The precise description of the energy budget at the sea ice cover, involving incoming and outgoing radiations
and the thermodynamics of ice, can be quite a challenging task \cite{MaykutUntersteiner,EbertCurry,EisenmanWettlaufer}.
Being the focus of our study, though, a particular aspect of the melting process, namely the reduced albedo by
meltwater covering the sea ice surface, we adopt a simple modelling \cite{Luethje}, that proves, on
the other hand, to be suitable to straightforward generalizations for the problems of interest here.
We write the total melting rate $f$ appearing in (\ref{eq:SM1}) as the sum of two terms
\begin{equation}\label{eq:f}
f = (1-\chi)\phi_1(w) + \chi \phi_2(w,\nabla w,\nabla h);
\end{equation}
here, the first term, $\phi_1$, is {\it local}, in fact it depends only on the pond depth $w(\mathbf{x},t)$, whereas
the second term, $\phi_2$, includes also {\it lateral melting} mechanisms and may, thus, in principle depend also
on gradients of the pond depth and ice thickness fields. The binary variable
$\chi \in \{0,1\}$ has been introduced to switch on ($\chi=1$) or off ($\chi=0$) such lateral melting contribution.
Let us first discuss the local term $\phi_1$.
We assume a constant melting rate $\phi_1 = m_i$, of dimensions $[\mbox{length}/\mbox{time}]$,
for {\it bare} (unponded)
ice (i.e. if $w(\mathbf{x},t)=0$), which 
is magnified by a $w$-dependent factor $\mathcal{A}(w)$, if  ice is covered by a pond ($w(\mathbf{x},t)>0$); altogether,
the expression for $\phi_1$ reads:
\begin{equation}\label{eq:mrluethje1}
\phi_1(w) = \mathcal{A}(w)m_i.
\end{equation}
Following L\"uthje et al. \cite{Luethje}, one can take $\mathcal{A}(w)$ to be:
\begin{equation} \label{eq:mrluethje2}
\mathcal{A}(w) =
\left\{
\begin{array}{ll}
1 + \frac{m_p}{m_i}\frac{w}{w_{\mbox{\tiny{max}}}} & \mbox{if } w \in [0,w_{\mbox{\tiny{max}}}] \\
& \\
1 + \frac{m_p}{m_i} & \mbox{otherwise}
\end{array}
\right. 
\end{equation}
where $m_p$ is a (constant) limit melting rate for ponded ice, when the overlying pond depth exceeds the value
$w_{\mbox{\tiny{max}}}$ (which is usually estimated to be pretty small,
  $w_{\mbox{\tiny{max}}} \approx 0.1 \m$, because turbulent convection is already relevant at such depth, as
  discussed later on). 
The meaning and origin of such magnifying factor deserves some comments.
In very shallow ponds, $w < w_{max}$, as a consequence of the absorption of solar radiation by water,
the warming up is proportional to its volume and so the heat flux through the liquid layer is proportional to $w$. 
The situation changes for slightly deeper ponds, $w > w_{max}$,  due to the appearance of natural convection. 
Indeed in summertime the temperature of air in contact with ponds ($\approx 2^o\,C$) is higher than the basal one,
in contact with melting ice (at $0^o\,C$). In this range water density shows the
well known anomaly, according to which it decreases with temperature, $\rho_w(T=2^0\,C) > \rho_w(T=0^0\,C)$,
therefore, the pond is prone to convection.
The latter sets on when the system becomes dynamically unstable; this will occur when the pond depth,
which grows in time because of melting
(thus making the system intrinsically non-stationary), will reach a value such that the time-dependent Rayleigh number
$Ra(t)$ is large enough. The Rayleigh number quantifies the relative magnitude of buoyancy and dissipative
terms; grouping together water density $\rho_w$, thermal expansion coefficient $\beta$, dynamic
viscosity $\eta$, thermal conductivity $\kappa$ and specific heat capacity at constant pressure $c_p$ with gravity yields:
\begin{equation} \label{eq:Ra}
  Ra(t) = \frac{c_p \rho_w^2 \beta g (\Delta T) w(t)^3}{\kappa \eta},
\end{equation}
Although it may seem surprising, the ponds being in general shallow, if we plug
typical values in (\ref{eq:Ra}) we get, even for $w \approx 0.1 \m$ and $\Delta T \approx 0.2^o\,C$,
$Ra \approx 10^6$ \cite{TaylorFeltham},
a value at which convection is already moderately
turbulent \cite{Ahlers}. Within ponds of depth $w \stackrel{>}{\sim}0.1 \m$, filled of fresh water, heat is not transferred
by conduction, but by turbulent convection, whence the larger basal melting rate
(\ref{eq:mrluethje1})-(\ref{eq:mrluethje2}). For the sake of simplicity we neglect here salt concentration.
  Such an assumption must be taken with due care, though, since salinity hinders convection, by density stratification,
  and can even inhibit it (as shown in \cite{Kim}). \\
The dependence of the turbulent heat flux $\Phi_{\mbox{\tiny{turb}}}$ (in $W/\textrm{m}^{-2}$ units) on the depth,
though, is a complex problem.
Expressed in non-dimensional variables, it amounts to assessing the Nusselt $Nu$ \textit{vs} Rayleigh numbers
scaling $Nu \sim Ra^{c}$ \cite{Ahlers,GL}, where the Nusselt number is defined as:
\begin{equation}
Nu(t) = \frac{\Phi_{\mbox{\tiny{turb}}}(t)}{\kappa \frac{(\Delta T)}{w(t)}}.
\end{equation}
The expression (\ref{eq:mrluethje2}) arises from the assumption of the so called Malkus scaling
$Nu \sim Ra^{1/3}$ \cite{Malkus}. Note that this state corresponds the conjecture that the turbulent heat flux is
independent of the thickness of the liquid layer, and as consequence that the melt rate is fixed at a constant value $m_p$
as stated by (\ref{eq:mrluethje2}) in the model by L\"uthje et al. \cite{Luethje} or by Taylor \& Feltham
\cite{TaylorFeltham}.
However, theories,
experiments and numerical simulations tend to agree that, in the range of $Ra$ of relevance for melt pond convection,
the scaling exponent should
be $c<1/3$ (see, e.g., \cite{GL} and references therein); in particular, widely observed is $Nu \sim Ra^c$,
with $c \approx 2/7$. A similar scaling was observed, in numerical simulations, also for turbulent thermal
convection with phase transition, where a boundary evolves, driven by melting \cite{Enrico}, a setup which more
closely resembles what occurs inside a melt pond.
So, we propose to generalize Eqs.~(\ref{eq:mrluethje1})-(\ref{eq:mrluethje2}) for the local magnitude of melting
to a generic $Nu \sim Ra^{c}$ relation and we obtain:
\begin{equation} \label{eq:gmrate}
\phi_1(w) = m_i + m_p(w,c)\left(\frac{w}{w_{\mbox{\tiny{max}}}}\right)^{\alpha}, \quad \mbox{with}\ 
\alpha =
\left\{
\begin{array}{ll}
1 & \ \ \mbox{if } w \in [0,w_{\mbox{\tiny{max}}}] \\
\\
3c - 1 &\ \  \mbox{if } w > w_{\mbox{\tiny{max}}},
\end{array}
\right. 
\end{equation}
so that for ponds deeper than $w_{max}$ L\"uthje et al.'s case \cite{Luethje} is recovered for $\alpha = 0$,
while scaling exponent equal to $2/7$ yields for $\alpha = -1/7$.  
Notice that we have allowed also the constant $m_p$ to be depth dependent in our model, $m_p \rightarrow m_p(w)$.
This is done in order to include another aspect of realistic convection in Arctic ponds:
the effect of a surface wind shear.
At high latitudes, in fact, strong wind shear from the atmospheric boundary layer is present that can affect
significantly sea ice dynamics (e.g. in the formation of sea-ice bridges \cite{Rallabandi}). Artic winds act
on pond surfaces and are able, in principle, to strongly modify the convection patterns \cite{SkyllingstadPaulson}.
In such situations, turbulent heat flux
is initially depleted, due to thermal plumes distortion by the shear \cite{Domaradzki,Scagliarini},
and then it increases again, when turbulent
forced convection becomes the dominant mechanism. 
On the line of the same arguments exposed in \cite{Scagliarini}, based on Prandtl's mixing length theory \cite{Prandtl},
an expression for the coefficient $m_p(w)$ of the following form 
\begin{equation} \label{eq:mprate}
  m_p(w,c) \sim  m_p^{(0)}(w,c)\left( \frac{a_1}{1 + c_1(\tau_s) w^{\gamma_1}} + a_2 c_2(\tau_s) w^{\gamma_2} \right),
\end{equation}
can be expected, where $a_1$ and $a_2$ are some phenomenological parameters and $c_1$ and $c_2$ are
functions of the wind shear magnitude $\tau_s$ (and of physical properties of meltwater).
In all numerical results reported here, however, we have set $\tau_s=0$,
  that is we have kept $m_p(w) \equiv m_p^{(0)}(w,c)$
  (exploring wind shear effects will be object of a forthcoming study). The dependence of $m_p^{(0)}(w,c)$
  on $w$ and $c$ stems from the fact that: i) below $w_{max}$ the heating is mainly radiative and ii)
  changing the exponent of the scaling relation between dimensionless
  quantities, $Nu \sim Ra^c$, affects also the prefactor of the turbulent heat flux, i.e.
  $\Phi_{\mbox{\tiny{turb}}} = A(c)w^{3c-1}$. The expression for $m_p^{(0)}(w,c)$ therefore reads:
\begin{equation} \label{eq:mprate0}
  m_p^{(0)}(w,c) =
\left\{
\begin{array}{ll}
m_{p,r}^{(0)} & \ \ \mbox{if } w \in [0,w_{\mbox{\tiny{max}}}] \\
\\
b_c(Pr) \left(\frac{c_p \rho_w^2 \beta g}{\eta}\right)^c \kappa^{1-c} (\Delta T)^{1+c} &\ \  \mbox{if } w > w_{\mbox{\tiny{max}}},
\end{array}
\right. 
\end{equation}
where the coefficient $b_c(Pr)$ depends on the Prandtl number, $Pr=c_p \eta/\kappa$.
As previously commented, Eq.~(\ref{eq:gmrate}) is purely local and ``vertical'', in the sense that,
if we think in discrete time, in a step $\Delta t$, it would increase the pond depth by
$\phi_1(w(\mathbf{x},t))\Delta t$,
$w(\mathbf{x},t) \rightarrow w(\mathbf{x},t)  + \frac{\rho_i}{\rho_w}\phi_1(w(\mathbf{x},t),t)\Delta t$,
and decrease the ice thickness by $h(\mathbf{x},t) \rightarrow h(\mathbf{x},t) - \phi_1(w(\mathbf{x},t))\Delta t$,
without affecting or being affected by the neighbourhood.
We may expect, though, that, due to convection induced mixing, meltwater will be at a higher temperature than
the surrounding ice and it may, therefore, favour melting also horizontally. This can be especially relevant close
to the edge of pond surfaces, where it should give rise to a widening of ponds.
To account for this kind of mechanism, we have introduced in the expression for the total melting rate,
Eq.~(\ref{eq:f}), the term $\phi_2(w,\nabla w, \nabla h)$, which contains the lateral melting 
(its explicit lattice expression will be given in section \ref{subsec:numerics}).
An attempt to estimate lateral fluxes in pond convection was proposed by Skyllingstad \& Paulson \cite{SkyllingstadPaulson},
though with prescribed and fixed (with no evolving boundaries) forms of ponds. 
Finally, it is important to underline that by ``lateral melting'' we refer here to horizontal melting within the pond,
and not edge melting of the ice pack, as when interactions with the ocean are considered \cite{Tsamados2015}.


\subsubsection{Seepage rate}

Sea ice has a complex porous structure that evolve in time as the pack melts \cite{FelthamMushy,Eicken2002};
a thorough description of water percolation through it is a formidable task that goes beyond the scope
of the present work. We just model water transport through sea ice using Darcy's law; in addition, we
distinguish between vertical and horizontal transport \cite{Luethje,ScottFeltham}.
Vertical transport is accounted for in Eqs.~(\ref{eq:SM1}) by the seepage term $s$; the horizontal
contribution, also dubbed lateral drainage, will be discussed in the next subsection.
In order to derive an expression for the seepage rate, we recall that, according to Darcy's law,
  the discharge through of homogeneous porous material of permeability $k$, cross-sectional area
  $a$ and length $\ell$, under an applied pressure difference $(p_{\mbox{\tiny{in}}} - p_{\mbox{\tiny{out}}})$,
  is given by
  \begin{equation}
    q = k \frac{a(p_{\mbox{\tiny{in}}} - p_{\mbox{\tiny{out}}})}{\eta \ell};
  \end{equation}
    for a portion of ponded ice of elementary area $\delta a$ and thickness $h$, such pressure head is due to the hydrostatic
  pressure of the column of water in the pond overlying ice on $\delta a$, whose height is $w$, is
  $(p_{\mbox{\tiny{in}}} - p_{\mbox{\tiny{out}}}) = \rho_w g \delta a w$. The discharge $q$ equals the time variation of
  the overlying volume of water, $\dot{\mathcal{V}} = \delta a \dot{w}$, providing 
  \begin{equation}
\dot{w} = -k\frac{\rho_w g w}{\eta h},
  \end{equation}  
  out of which we can read the expression for the seepage rate $s$ that is \cite{ScottFeltham}
\begin{equation}\label{eq:seepage}
 s = k\frac{\rho_w g}{\eta}\frac{w}{h}.
\end{equation}  

\subsubsection{Meltwater flux}

The seepage rate just introduced, Eq.~(\ref{eq:seepage}), entails a dependence of the equation for
$w(\mathbf{x},t)$ on $h(\mathbf{x},t)$ (that would be otherwise be decoupled from it, as far as only
melting is concerned). A further coupling is induced by the transport term and the associated
meltwater flux $\mathbf{u}$.
Such term is also the only non-local one in the evolution (for it involves derivatives of $h$ and $w$), thus
introducing a dependence of the dynamics on the ice topography. It represents, in other words, the driving for
meltwater to accumulate to form ponds.
The transport of meltwater is realised essentially with two mechanisms: {\it sliding} of water over slopes of
the ice surface and {\it lateral drainage} through the porous structure of ice. Correspondingly, the flux consists
of the sum of two terms
\begin{equation}
\mathbf{u} = \mathbf{u}_{\mbox{\tiny{sliding}}} + \mathbf{u}_{\mbox{\tiny{drainage}}}; 
\end{equation}
as discussed in the previous subsection, $\mathbf{u}_{\mbox{\tiny{drainage}}}$ stems from the horizontal
component of Darcy's law and, hence, is given by \cite{Luethje}
\begin{equation}\label{eq:latdrainage}
\mathbf{u}_{\mbox{\tiny{drainage}}} = -\Pi\frac{\rho_w g}{\eta} \nabla (h + w), 
\end{equation}  
where $\Pi$ is the horizontal permeability of ice. \\
In order to model the sliding term, we resort to the theory of shallow water equations (SWE) \cite{Landau},
considering that the width of a layer of water sliding over the ice topography is relatively thin.
If we assume, furthermore, that the Reynolds number is small (we expect so, and a consequent creeping
  flow, for a thin layer of water sliding over the ice topography, the thickening of such layer being
  inhibited by seepage),
the SWE for the depth-averaged two-dimensional
velocity field reduce to the following balance equation between stresses at the bottom (due to friction with ice) and
top (induced by wind forcing) of the fluid layer and gravity \cite{Marche,Oron}
(assuming a no-slip boundary condition between water and ice and neglecting capillary effects)
\begin{equation}
\frac{3 \eta}{w} \mathbf{u}_{\mbox{\tiny{sliding}}} + \mathbf{\tau}_s +g w \nabla (h + w) \approx 0, 
\end{equation}
which yields for $\mathbf{u}_{\mbox{\tiny{sliding}}}$:
\begin{equation}\label{eq:sliding}
\mathbf{u}_{\mbox{\tiny{sliding}}} = - \frac{g w^2}{3 \eta} \nabla(h+w) + \frac{\tau_s w}{3\eta} \hat{\tau}_s, 
\end{equation}
where $\hat{\tau}_s$ is the direction of the wind shear vector at the free water surface and
$\tau_s$ is its magnitude, as in Eq.~(\ref{eq:mprate}).
Let us stress that, in this way, we have introduced, through Eqs.~(\ref{eq:mprate}) and (\ref{eq:sliding})
a first minimal coupling of the model for the sea-ice-melt-ponds system with the atmospheric dynamics.

\subsection{Numerical implementation} \label{subsec:numerics}

The system of equations (\ref{eq:SM1}) is solved by means of a
finite differences scheme; upon discretization on a square $M \times M$ lattice, with $M=1024$,
of equally $\Delta$-spaced nodes, the system is converted in a set of
coupled ordinary differential equations for the variables $h_{ij}(t) \equiv h(x_i,y_j,t)$ (with $x_i = i\Delta$,
$y_j = j\Delta$ and $i,j=1,2,\dots,M$) and
$w_{ij}(t) \equiv w(x_i,y_j,t)$, that are, then, integrated numerically using a standard explicit Runge-Kutta
$4$th order time marching scheme with time step $\Delta t = 60 \mbox{s}$, that allows to resolve
  the fastest time scales of the meltwater transport terms.
Spatial derivatives are approximated by the corresponding second order accuracy central differences.
The lattice spacing $\Delta$ is taken to be $\Delta = 1 \m$, so the
physical size of the simulated system is $L^2 \approx 1 \mbox{km}^2$, where $L=M\Delta$;
this choice is dictated by the condition that $\Delta$ is $\Delta \stackrel{>}{\sim} \sigma_h$
  ($\sigma_h$ being the
  standard deviation of the initial ice thickness distribution), such that no significant variations of
  of $h$ occur within one lattice spacing, i.e. the spatial derivative is at most $h^{\prime}(x) \sim 1$,
assuming that the average finite height variation over a $\Delta$ is $\Delta h \propto \sigma_h$.
Periodic boundary conditions apply, so we
neglect edge effects, such as water run-off and direct coupling with the ocean (e.g. lateral melting of floe,
ocean stresses), i.e. it is as if we were simulating a virtually infinite sea ice floe.\\
The melting term $\phi_2$, appearing in eq. (\ref{eq:f}), takes the following expression on the lattice
\begin{equation}\label{eq:latmelt}
\phi_{2_{i,j}} = \phi^{(V)}_{2_{i,j}} + \sum_{i^{\prime}=\pm 1}\phi^{(L,x)}_{2_{i+i^{\prime},j}}
  \Theta(w_{i+i^{\prime},j} - w_{i,j})+ \sum_{j^{\prime}=\pm 1}\phi^{(L,y)}_{2_{i,j+j^{\prime}}}
  \Theta(w_{i,j+j^{\prime}} - w_{i,j}),
\end{equation}   
which contains a combination of {\it vertical}, $\phi^{(V)}_{2_{i,j}}$, and {\it lateral}, $\phi^{(L,(x,y))}_{2{i,j}}$,
components of the melting; the latter are given by: 
\begin{equation}\label{eq:vertmelt}
  \phi^{(V)}_{2_{i,j}} = \phi_{1_{i,j}} \frac{1}{\sqrt{1 + (\hat{\partial}_x w_{i,j})^2 + (\hat{\partial}_y w_{i,j})^2}}
\end{equation}  
and
\begin{equation}\label{eq:latlatmelt}
  \phi^{(L,(x,y))}_{2,{i,j}} = \phi_{1_{i,j}} \frac{|\hat{\partial}_{(x,y)}w_{i,j}|}{\sqrt{1 + (\hat{\partial}_x w_{i,j})^2 + (\hat{\partial}_y w_{i,j})^2}},
\end{equation}  
where $\hat{\partial}_{(x,y)}$ stands for the finite difference derivative.
We assume that the magnitude of the turbulent heat flux
is homogeneously distributed over the pond walls (that is at the ice/water interface) and its direction is parallel
to the normal $\hat{n}$ to the interface. Therefore, the vertical and lateral contributions to the melting rate are
weighted with the absolute values of the components of $\hat{n}$,
$$
\frac{1}{\sqrt{1 + (\hat{\partial}_x w_{i,j})^2 + (\hat{\partial}_y w_{i,j})^2}}
\left(|\partial_x w|,|\partial_y w|,1\right),
$$
whence eqs.~(\ref{eq:vertmelt}) and (\ref{eq:latlatmelt}).
In other words,
this means that, for instance,
at the bottom of the pond mostly the vertical term will act, while when the topography is steep, as, e.g., next to the
pond edge, ice ablation will be dominated by lateral melting.
The presence of the Heaviside's functions, $\Theta$, in (\ref{eq:latmelt}) is to guarantee that the,
non-local, lateral
contribution to melting on a given site comes only from those neighbours that
have a larger amount of overlying water (larger $w$).
  This is motivated by the idea that, if at a given elevation $H$ a certain site is in the 'ice state',
  it will get a lateral melting contribution from a neighbouring site which, at the same elevation, is in a
  'water state', since melting is driven by water convection in contact with ice enclosing the pond.

\section{Results} \label{sec:results}

The initial values of the sea ice topography $h^0_{ij} \equiv h(x_i,y_j,0)$ are random Gaussian numbers with given
mean and variance. The initial topography is spatially correlated over a characteristic length
$\delta \approx 8 \, \m$.
Two types of ice are used as initial conditions, namely first-year ice (FYI) and multi-year ice (MYI). FYI is newly
formed in the winter preceding the melt season and is typically flatter, whereas MYI, that
has overcome one or more melt seasons, presents a more rugged surface profile, i.e. it is characterized by larger
variance and mean as compared to FYI. Consequently, wide and ramified but shallow melt ponds are more probably formed
on FYI, while melt ponds on MYI will be tendentially deeper, of limited areal extension and more regularly shaped
\cite{FloccoFeltham}. The initial condition is therefore expected to play an important role on the meltwater dynamics.
The statistical parameters (mean $\langle h \rangle$ and variance $\sigma_h$ of the thickness distribution) employed
are $\langle h \rangle = 0.92 \,\m$, $\sigma_h = 0.18 \,\m$, for FYI,
and $\langle h \rangle = 3.67 \,\m$, $\sigma_h = 1.5 \,\m$, for MYI \cite{Luethje,Hvidegaard}.
Other numerical values for the model parameters, which are kept fixed in all simulations, are summarized
in Table \ref{tab:params}. Evidently, we are faced to a wide, multi-dimensional, parameter space;
  many of these parameters (such as permeabilities and melting rates) are known only with limited accuracy and
  the system can be quite sensitive to their values. A full sensitivity study in such sense is somehow beyond the
  scope of the present work; moreover some studies of this kind (on similar models) are available
  (see, e.g. \cite{Luethje,ScottFeltham}). We limit here ourselves, therefore, to test the novelties of the present
  model, namely the melting rate exponent associated to turbulent thermal convection and its contribution along the lateral
  (horizontal) directions.

\begin{table}
  \caption{Values of model parameters which are kept fixed in all simulations: water density $\rho_w$, ice density $\rho_i$, water dynamic
    viscosity $\eta$, acceleration of gravity $g$, horizontal permeability of ice $\Pi$, {\it bare} ice melting rate $m_i$, melting rate
    enhancement factor $m_p^{(0)}$ and critical pond depth for melting rate enhancement $w_{\mbox{\tiny{max}}}$.}
\label{tab:params}       

\begin{tabular}{p{1.5cm}|p{1.cm}p{1.cm}p{1.6cm}p{1.cm}p{1.3cm}p{1.1cm}p{1.1cm}p{1.cm}}
\hline\noalign{\smallskip}
Parameter & $\ \rho_w $ & $\rho_i $ & $\eta $& $g $ & $\Pi $ & $m_i $ & $m_p^{(0)} $ & $w_{\mbox{\tiny{max}}} $  \\
\noalign{\smallskip}\svhline\noalign{\smallskip}
Units & $\ \mbox{kg}/\m^3$ & $\mbox{kg}/\m^3$ & $\mbox{kg}/(\m \; \mbox{s}^{-1})$ & $\m/\mbox{s}^2$ & $\m^2$ &$\mbox{cm}/\mbox{day}$ & $\mbox{cm}/\mbox{day}$ & \m\\
\noalign{\smallskip}\svhline\noalign{\smallskip}
Value & $\ 1000$  & $950$  & $1.79 \times 10^{-3}$ & $9.81$ & $3 \times 10^{-9}$ & $1.2$ & $2$ & $0.1$ \\
\noalign{\smallskip}\hline\noalign{\smallskip}
\end{tabular}
\end{table}

\begin{figure}[b]
\sidecaption
\includegraphics[scale=1.0]{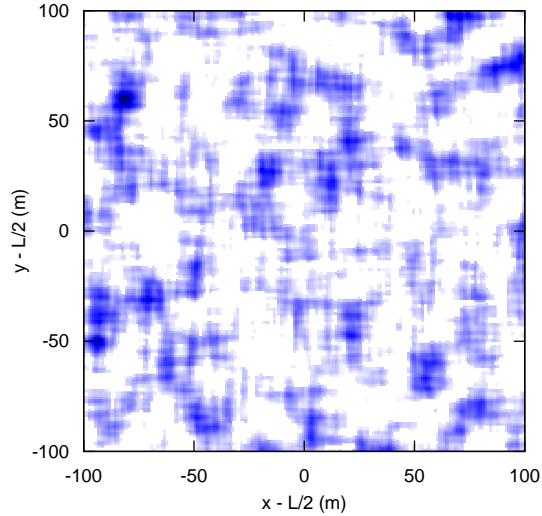}
\caption{Configuration of the depth field $w(\mathbf{x},t)$ showing the melt ponds distribution over the sea ice surface, for FYI after 20 simulated days (a $200 \times 200 \m^2$ region at the centre of the simulated domain is taken).
  White color corresponds to bare ice and blue color indicates the presence of a pond, the darker the blue the deeper
the pond (deepest ponds have $w \approx 2 \m$).}
\label{fig:snap}
\end{figure}

Snow cover is absent and no melt water is assumed at the initial time
(i.e. $w(\mathbf{x},t_0)=0 \quad \forall \mathbf{x}$).
As said before, we aim to simulate the summer time evolution of sea ice, so our $t_0$ is to be considered June 1st and,
in view of this, refreezing of meltwater is not accounted for. We ran each simulation for $\approx 30$ days. 
A visualization of the distribution of ponds corresponding to day $20$ from the beginning of the simulation is shown
in figure \ref{fig:snap}
In order to extract statistical informations on the melt pond coverage of the sea ice, we first need to identify
individual ponds. To do this, for each time $t$ we define a pond as any connected subset of points on the
lattice such that $w(\mathbf{x},t)>0$; the full pond configuration is determined by a cluster analysis (for which
we employ the so called Hoshen-Kopelman algorithm \cite{HK}) over the whole system. The area of the $i$-th pond is
then $A_i = n_i \Delta_x \Delta_y$, $n_i$ being the number of points in the $i$-th cluster.

\subsection{Melt pond areas evolution: role of the turbulent heat flux scaling inside the pond}

\begin{figure}[b]
\sidecaption
\includegraphics[scale=.8]{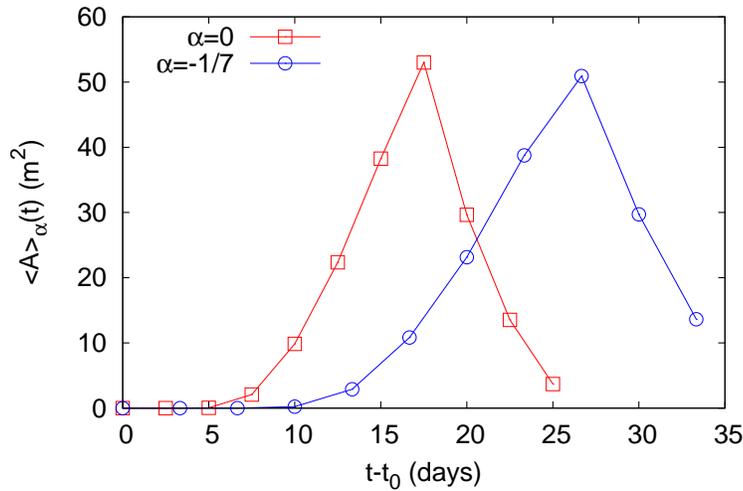}
\caption{Mean pond area vs time for the $1/3$ (red squares) and $2/7$ (blue circles) laws.}
\label{fig:amean}
\end{figure}

\begin{figure}[b]
\sidecaption
\includegraphics[scale=.8]{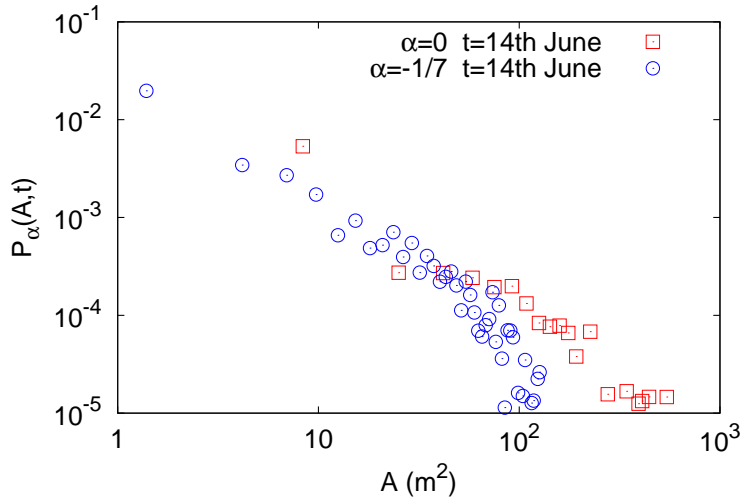}
\caption{Probability distribution functions of pond areas for the $1/3$ and $2/7$ laws at $14$th June.}
\label{fig:pdfs-eqtime} 
\end{figure}

\begin{figure}[b]
\sidecaption
\includegraphics[scale=.8]{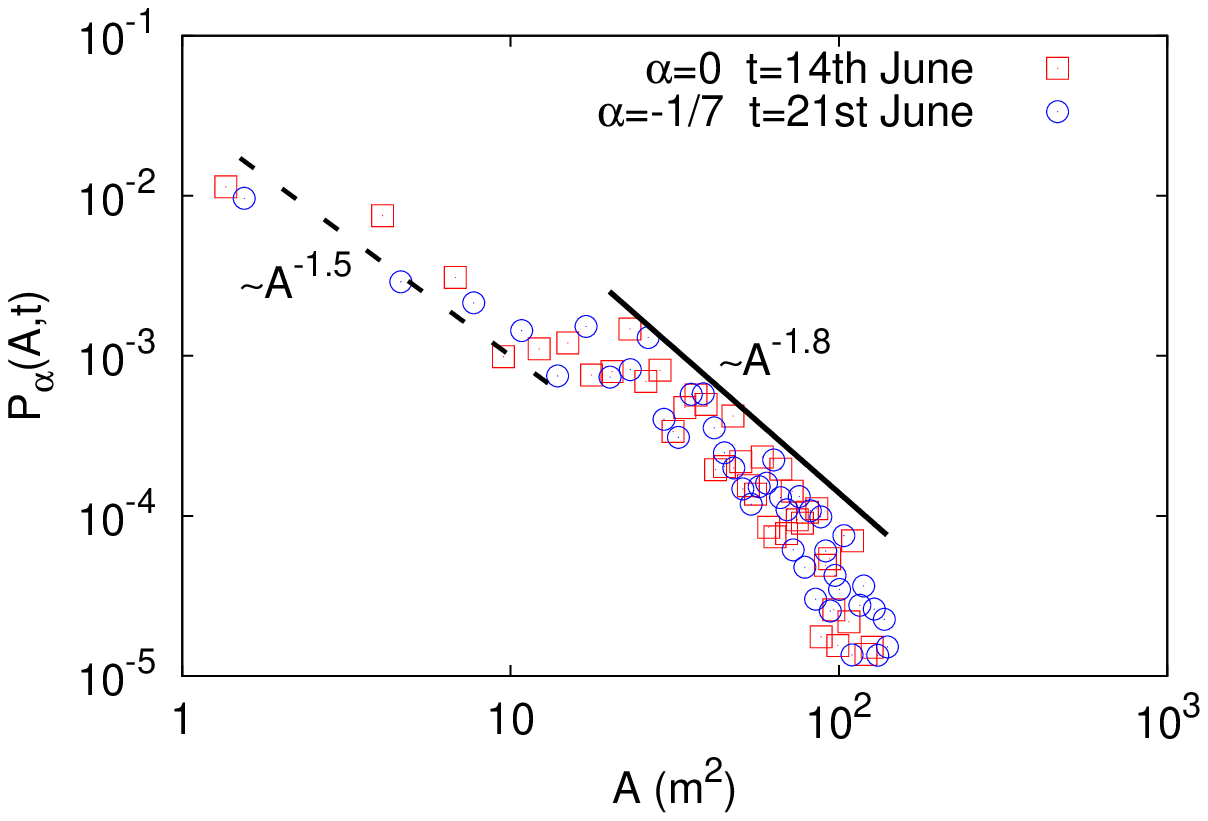}
\caption{PDFs of pond areas after $13$, for $\alpha=0$, and $20$ days, for $\alpha=-1/7$: notice that the two
  sets of points basically overlap. The dashed and solid lines correspond to the power law $A^{1.5}$ and $A^{1.8}$,
respectively.}
\label{fig:pdfs-eqmean}
\end{figure}

In figure \ref{fig:amean} we plot the time evolution of the mean pond area
\begin{equation}
\langle A \rangle_{\alpha}(t) = \frac{1}{N(t)}\sum_{i=1}^{N(t)} A_i(t)
\end{equation}
(where $N(t)$ is the total of ponds detected at time $t$) for a FYI and assuming Malkus and $2/7$ scaling for
the turbulent heat flux, respectively, that is, with reference to Eq.~(\ref{eq:gmrate}), with
$\alpha=0$ (red squares, equivalent to the study in \cite{Luethje}) and $\alpha=-1/7$ (blue circles).
The mean pond area grows and reaches a maximum faster when $\alpha=0$: after $13$ days, e.g., the $2/7$-model
gives a prediction for $\langle A \rangle_{\alpha}$ approximately seven times
smaller than it is for the constant flux case; 
 this suggests how an apparently
minor assumption at the level of fluid dynamic processes within the single pond may lead to bad estimates on
climatologically relevant indicators, such as the September sea ice extension. 
For the same two runs, with $\alpha=0,-1/7$, we measured the probability distribution functions (PDFs) of pond areas,
$P_{\alpha}(A,t)$, after $13$ days;
one can see from figure \ref{fig:pdfs-eqtime} that the two PDFs differ, although both seem to show
a power law behaviour. 
Nevertheless, if we consider PDFs with {\it equal mean}, instead of {\it equal time} PDFs, interestingly, the two sets
of points (for $\alpha=0$ and $\alpha = -1/7$) collapse onto each other, as shown in figure
\ref{fig:pdfs-eqmean}. There we plot $P_{\alpha=0}(A,t_1)$ and $P_{\alpha=-1/7}(A,t_2)$, where $t_1$ and $t_2$
are such that $\langle A\rangle_{\alpha=0} (t_1) = \langle A\rangle_{\alpha=-1/7} (t_2)$; with reference to
figure \ref{fig:amean}, this occurs, for instance, if we pick $t_1 - t_0 = 13$ days and $t_2 - t_0 = 20$
  days, i.e. on June $14$th for $\alpha=0$ and June $21$st for $\alpha=-1/7$. 
The two PDFs nicely follow the scaling $P_{\alpha}(A) \sim A^{1.5}$ for relatively small areas
($A < 20 \m^2$), with a steeper fall-off for larger values, $P_{\alpha}(A) \sim A^{1.8}$.  
Such functional forms agree with available observational data, as those collected by means of aerial photography
\cite{Perovich2002aerial} during the SHEBA (Surface Heat Budget of the Arctic Ocean) \cite{MoritzEtAl,MoritzPerovich}
and HOTRAX ({\it Healy–Oden} TRans Arctic EXpedition) \cite{PerovichEtAl2009} campaigns.
Remarkably, the same power-laws for the PDFs were found in recent theoretical/numerical works
based on statical models (in the spirit of equilibrium statistical mechanics) \cite{Ising,Popovic}.
We would like to highlight, at this point, that this is a striking aspect of the melt-pond-sea-ice system:
the melt pond system on large scales
is robust with respect to area distribution (and pond geometry, as we shall see later on) against changes of
certain physical parameters controlling the dynamics. So robust that even simple models, that do not account for
the physics of the melt pond formation and evolution at all, can capture such statistical fingeprints.
We will focus, then, on an aspect of melt pond configuration that one might expect to be affected
by details of the evolution, namely their morphology.  

\subsection{Morphology of melt ponds: role of lateral melting}

Characterizing the morphology of the global ponds configuration and understanding how it emerges
can be of great relevance also for large scale models of sea ice (in GCMs). There, in fact, a major limitation
is due to the difficulty to relate properly the sea ice topography with the redistribution of meltwater; ideally,
one would wish to know how much ice area is covered by water, and how deeply (since these two quantities determine,
basically, the absorbance of incident radiation).
\begin{figure}[h]
\sidecaption
\includegraphics[scale=.8]{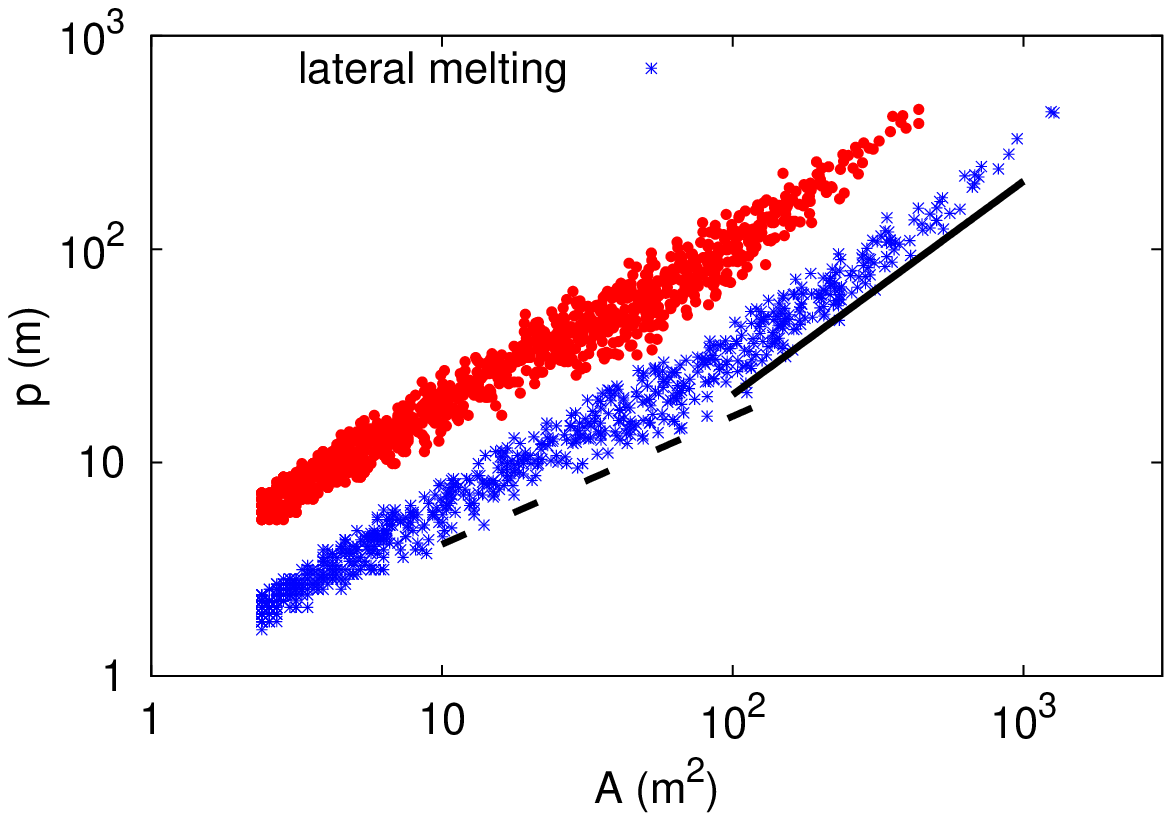}
\caption{Scatter plot of perimeter and area for all melt ponds on $21$st June. The dashed and solid lines indicate
  the power law $A^{1/2}$ and $A$, corresponding to perimeter fractal dimensions $d_p=1$ (smooth shapes) and
  $d_p=2$ (fractal), respectively; the transition between the two regimes occur at $A \approx 100 \m^2$.}
\label{fig:PvsA}
\end{figure}
Analyzing aerial images from two different Arctic expeditions, SHEBA \cite{MoritzEtAl,MoritzPerovich}
and HOTRAX \cite{PerovichEtAl2009}, Hohenegger and coworkers
\cite{Hohenegger} looked at the scatter plot of perimeter $p$ and area $A$ of a multi-pond configuration; such a plot
is known to contain informations on the fractal geometry of the manifold (embedded in a two-dimensional space)
considered \cite{Mandelbrot,ChengMatGeo}.
The two quantities are, in fact, related by
\begin{equation}\label{eq:PvsA}
p \sim A^{d_p/2},
\end{equation}  
where $d_p$ is the so called perimeter fractal dimension: for a smooth curve $d_p = 1$, while for a fractal,
in the strict sense, $d_p>1$. It was observed that surfaces of small ponds tend to be of roundish
shape, while large
ones, that typically stem from aggregation of several small ponds, display features of clusters in percolating
systems and appear fractal-like \cite{Hohenegger}.
This transition to a fractal geometry is supposedly connected with the way melt ponds grow
over the sea ice surface; it is natural to ask, then, whether the explicit modelling of the physical mechanisms
leading to such in-plane growth has any impact on the final global morphology.
This amounts to test the model in presence of what we called lateral melting, i.e. with
a melting rate given by Eqs.~(\ref{eq:f}), with $\chi=1$, and (\ref{eq:latmelt}).
\begin{figure}[h]
\sidecaption
\includegraphics[scale=.8]{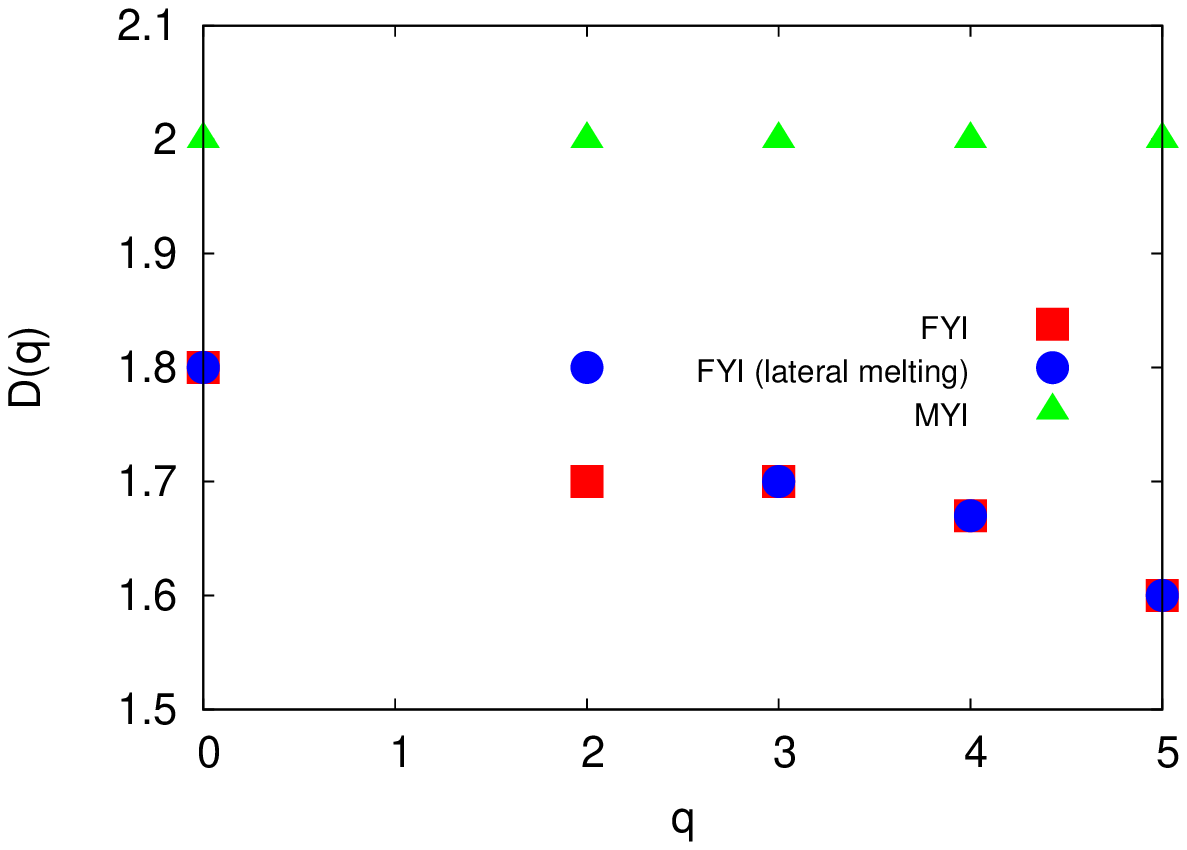}
\caption{Generalized fractal (or Renyi's) dimensions for the melt pond distribution after $70$ days from three
  simulations: FYI without lateral melting (red squares), FYI without lateral melting (blue dots) and MYI 
  without lateral melting (green triangles).}
\label{fig:Dq}
\end{figure}
To this aim, we ran the same simulation, for the FYI, as discussed above, with the lateral melting term
switched on. In figure \ref{fig:PvsA} we show $p$ vs $A$ scatter plot after $20$ days for two simulations with
(blue asterisks) and without (red bullets) lateral melting
(symbols relative to the two data sets are shifted from each other by a factor $3$ for the sake of clarity,
otherwise they would overlap).
The two power laws, $p \sim A^{1/2}$, for $A < A_c$, and $p \sim A$, for $A>A_c$, are reasonably well
followed in both cases; the only minor effect of the presence of lateral melting seems to be a
slightly clearer scaling behaviour (especially for large $A$). The transition to the fractal geometry occurs
at $A_c \approx 100 \m^2$, in agreement with the observations \cite{Hohenegger}. 
This same phenomenology was captured also by the above mentioned statistical models \cite{Ising,Popovic},
underlining further the robustness of the melt-pond-covered sea ice system as far as geometry is concerned. \\
To get a deeper insight on this aspect of melt pond configuration over the sea ice surface, we performed
an analysis of the {\it generalized fractal dimensions} (GFD), or R\'enyi's $q$-entropies,
$D(q)$ \cite{Grassberger}, which
provide a more detailed description of the geometry of the fractal manifold.
Let us call $\mu_i(\varepsilon)$ the measure of
ponds within the $i$-th element of a regular tessellation of the domain in squares of side $\varepsilon$
(i.e. the fractional area of the $\varepsilon$-square occupied by meltwater), and $N(\varepsilon)$ the total number of
squares into which the domain is partitioned; we can then define the following quantity:
\begin{equation}\label{eq:Iqe}
  I(q,\varepsilon) = \frac{1}{1-q}\log \left(\sum_{i=1}^{N(\varepsilon)} \mu_i(\varepsilon)^q\right),
\end{equation}
for any positive $q \neq 1$. The GFD are then computed as
\begin{equation}\label{eq:Dq}
  D(q) = \lim_{\varepsilon \rightarrow 0} \frac{I(q,\varepsilon)}{\log(1/\varepsilon)};
\end{equation}
for $q=0$, $I(0,\varepsilon)$ equals the number of non-void elements of the tessellation, therefore $D(q=0) \equiv D_0$
coincides with the Haussdorf, or box-counting, dimension, which is an estimate of the fractal dimension of the set
\cite{Mandelbrot}. It is clear, then, in which sense the $D(q)$ are generalized fractal dimensions.
For ``ordinary'' fractals, all the GFD are equal, i.e. $D(q)$ is constant with $q$. In general, though, it might
be a non-increasing function of the order $q$: if this is the case, one talks about a {\it multifractal} set, that is
a fractal whose dimension vary in space. 
In figure \ref{fig:Dq} we show the $D(q)$ computed from numerical data from three simulations, namely:
FYI without
lateral melting, FYI with lateral melting and MYI. The plot tells us that melt ponds on FYI lay on a fractal manifold,
as suggested also by the perimeter-area relation, since $D_0 < 2$, whereas those on MYI do not $D(q) = 2 \; \forall q$;
however, we observe a modest decrease of $D(q)$ with $q$, indicating a weak multifractality, with
very minor (if any) differences between the run with lateral melting and the one without.
These results are a first attempt to show that the morphology of the melt pond system can be even more complicated
than what can be captured with the perimeter-area relations, which are known to give sometimes biased
  estimates
of the actual fractal dimension for the areas \cite{ChengMatGeo}.

\section{Conclusions and perspectives}\label{sec:concl}

We have proposed a continuum {\it mesoscale} model that describes the evolution of Arctic sea ice, in presence
of a coverage of meltwater ponds, that alter the sea ice thermodynamics (in terms of melting rates).
The model consists of two coupled partial differential equations, for the ice thickness and pond depth fields.
The physics of sea ice was kept at a very basic level in order to focus on the effect of dynamic processes
occurring at the single pond level on the large scale configurations of the melt pond system and on sea ice evolution.
Numerical simulations of the model showed that a minimal variation of the scaling exponent of the turbulent heat
flux, within the pond, with the surface temperature impacts the time evolution of the mean pond size (shifting
the maximum by few days), hence of the average melting rate. We stress that the assumption made in this work, that the melt rate in ponds (deeper than $w_{max}$) is a weakly decreasing function of the water layer depth
rather than a constant is supported by a vast amount of studies on turbulent heat transfer.
Therefore, our study suggests that a thorough knowledge
and parametrization of melt pond hydrodynamics is needed in order to not get wrong estimates of observables of
climatological relevance. 
On the other hand, statistical and geometrical properties of the melt pond system,
such as the probability distribution function of pond surface areas and fractal dimensions, appeared to be robust
against heat flux scaling variations as well as against the inclusion or not of an explicit modelling of lateral
melting inside the pond.
In particular, our results agreed well with observation for what concerns the power law decay of the PDFs and the
perimeter-area relation, and the corresponding perimeter fractal dimension, for ponds.
Finally, we have extended the study of melt ponds geometry to the analysis of generalized fractal dimensions, which
showed a clear dependence on the initial ice topography, with melt ponds on first-year ice displaying even
a weak multifractality, while those on multi-year ice being essentially smooth.
This dependence on the initial condition suggests that, for future, studies, it would be of great interest
to initialize the numerical model with conditions taken by field measurements.
The study of the effect of wind stresses at the ice surface on global melting as well as on melt pond distribution and
morphology will be a first extension of the present study.
A further step forward that might be taken, within this approach, is the inclusion of a proper description of
mechanical processes and rheology of sea ice, specially focusing on the effect of the presence of accumulation
of meltwater on the local deformation properties of the pack.

\begin{acknowledgement}
  AS and DM acknowledge financial support from the National Group of Mathematical Physics of the Italian National
  Institute of High Mathematics (GNFM-INdAM).
  EC acknowledge supports form the French National Agency for Research (ANR) under the grant SEAS (ANR-13-JS09-0010).
\end{acknowledgement}
%

%
%

\begin{thebibliography}{99.}%

\bibitem{Hunke2010} Hunke, E.C., Lipscomb, W.H. and Turner, A.K.: Sea-ice models for climate study: retrospective and new directions,
  J. Glaciol. \textbf{56}, 1162--1172 (2010)

\bibitem{Notz2012} Notz, D.: Challenges in simulating sea ice in Earth System Models,
  WIREs Clim. Change \textbf{3}, 509--526 (2012)

\bibitem{CattleCrossley} Cattle, H. and Crossley, J.: Modelling Arctic climate change,
  Philos. Trans. Royal Soc. A 352, 1699 (1995)

\bibitem{Ebert1995} Ebert, E.E., Schramm, J.L. and Curry, J.A.:
  Disposition of solar radiation in sea ice and the upper ocean,
  J. Geophys. Res. 100, C8, 965--975 (1995)

\bibitem{MaykutMcPhee} Maykut, G.A. and McPhee, M.G.: Solar heating of the arctic mixed layer,
  J. Geophys. Res. 100(C12), 24,691--24,703 (1995)

\bibitem{Tsamados2014} Tsamados, M, Feltham, D.L., Schr\"oder, D., Flocco, D., Farrell, S.L., Kurtz, N., Laxon, S.W., and Bacon, S.:
  Impact of variable atmospheric and oceanic form drag on simulations of arctic sea ice,
  J. Phys. Oceanogr. 44, 1329--1353 (2014)

\bibitem{Vancoppenolle2009a} Vancoppenolle, M., Fichefet, T., Goosse, H., Bouillon, S., Madec, G.,
  and Morales Maqueda, M.A.:
  Simulating the mass balance and salinity of Arctic and Antarctic
  sea ice. 1. Model description and validation,
  Ocean Model. 27, 33--53 (2009)

\bibitem{Vancoppenolle2009b} Vancoppenolle, M., Fichefet, T. and Goosse, H.:
  Simulating the mass balance and salinity of Arctic and Antarctic
  sea ice. 2. Importance of sea ice salinity variations,
  Ocean Model. 27, 54--69 (2009)

\bibitem{Mauritzen} Mauritzen, C. and H\"akkinen, S.:
  Influence of sea ice on the thermohaline circulation in the Arctic-North Atlantic Ocean,
  Geophys. Res. Lett. 24, 3257--3260 (1997) 
  
\bibitem{Kwok2009} Kwok, R., Cunningham, G.F., Wensnahan, M., Rigor, I., Zwally, H.J. and Yi, D.: Thinning and volume loss of the Arctic Ocean sea ice cover: 2003--2008,
  J. Geophys. Res. 114, C07005 (2009)

\bibitem{Stroeve2012} Stroeve, J.C., Kattsov, V., Barrett, A., Serreze, M., Pavlova, T., Holland, M. and Meier, W.N.: Trends in Arctic sea ice extent from CMIP5, CMIP3 and observations,
  Geophys. Res. Lett. 39, L16502 (2012)

\bibitem{Laxon2013} Laxon, S.W., Giles, K.A., Ridout, A.L., Wingham, D.J., Willatt, R., Cullen, R., Kwok, R., Schweiger, A., Zhang, J., Haas, C.,
  Hendricks, S., Krishfield, R., Kurtz, N., Farrell, S. and Davidson, M.:
  CryoSat--2 estimates of Arctic sea ice thickness and volume, 
  Geophys. Res. Lett. 40, 732--737 (2013)

\bibitem{MaykutUntersteiner} Maykut, G.A. and Untersteiner, N.: Some results from a time-dependent thermodynamic model of sea ice,
  J. Geophys. Res. 76(6), 1550--1575 (1971)

\bibitem{EbertCurry} Ebert, E.E. and Curry, J.A.: An intermediate one--dimensional thermodynamic sea ice model for
  investigating ice--atmosphere interactions,
  J. Geophys. Res., 98(C6), 10,085--10,109 (1993)

\bibitem{EisenmanWettlaufer} Eisenman, I. and Wettlaufer, J.S.: Nonlinear threshold behavior during the loss of Arctic sea ice,
  Proc. Nat. Acad. Sci 106, 28--32 (2009)

\bibitem{Steele1992} Steele, M.: Sea ice melting and floe geometry in a simple ice--ocean model,
  J. Geophys. Res. 97(C11), 17, 729--17, 738 (1992)

\bibitem{BitzLipscomb} Bitz, C.M. and Lipscomb, W.H.: An energy-conserving thermodynamic model of sea ice,
  J. Geophys. Res. 104(C7), 15, 669–15, 677 (1999)
  
\bibitem{FreitagEicken} Freitag, J. and Eicken, H.: Melt water circulation and permeability
of Arctic summer sea ice derived from hydrological field experiments. J. Glaciol. \textbf{49}, 349--358 (2003)

\bibitem{FelthamMushy} Feltham, D.L., Untersteiner, N., Wettlaufer, J.S. and Worster, M.G.:
  Sea ice is a mushy layer, Geophys. Res. Lett. 33, L14501 (2006)

\bibitem{WellsMushy} Wells, A.J., Wettlaufer, J.S. and Orszag, S.A.:
  Nonlinear mushy-layer convection with chimneys: stability and optimal solute fluxes,
  J. Fluid Mech. 716, 203--227 (2013)
  
\bibitem{TurnerHunke} Turner, A.K. and Hunke, E.C.: Impacts of a mushy-layer thermodynamic 
  approach in global sea-ice simulations using the CICE sea-ice model,
  J. Geophys. Res. Oceans 120(2), 1253--1275 (2015)

\bibitem{FelthamRheo} Feltham, D.L.: Sea ice rheology, Annu. Rev. Fluid Mech. 40, 91--112 (2008)

\bibitem{HunkeDukowicz} Hunke, E.C. and Dukowicz, J.K.:
  An elastic-viscous-plastic model for sea ice dynamics,
  J. Phys. Oceanogr. 27(9), 1849--1867 (1997)

\bibitem{Tsamados2013} Tsamados, M., Feltham, D.L. and Wilchinsky, A.:
  Impact of a new anisotropic rheology on simulations of Arctic sea ice,
  J. Geophys. Res. Oceans 118(1), 91--107 (2013)

\bibitem{Rabatel2018} Rabatel, M., Rampal, P.,  Carrassi, A., Bertino, L. and Jones, C.K.R.T.: 
  Impact of rheology on probabilistic forecasts of sea ice trajectories: application for search and
  rescue operations in the Arctic,
  The Cryosphere 12, 935--953 (2018)
  
\bibitem{Steele1997} Steele, M., Zhang, J., Rothrock, D. and Stern, H.:
  The force balance of sea ice in a numerical model of the Arctic Ocean,
  J. Geophys. Res. Oceans 102(C9), 21,061--21,079 (1997)

\bibitem{Schroeder2003} Schr\"oder, D., Vihma T., Kerber, A. and Br\"ummer, B.:
  On the parameterization of turbulent surface fluxes over heterogeneous sea ice surfaces,
  J. Geophys. Res. 108(C6), 3195 (2003)

\bibitem{Rampal2009} Rampal, P., Weiss, J. and Marsan, D.:
  Positive trend in Arctic sea ice mean speed and deformation 1979-2007,
  J. Geophys. Res. Oceans 114, C05013 (2009)

\bibitem{Rampal2011} Rampal, P., Weiss, J., Dubois, C. and Campin, J.-M.:
  IPCC climate models do not capture the Arctic sea ice drift acceleration:
  Consequences in terms of projected sea ice thinning and decline,
  J. Geophys. Res. Oceans 116, C00D07 (2011)

\bibitem{Petty2013} Petty, A.A., Feltham, D.L. and Holland, P.R.:
  Impact of atmospheric forcing on Antarctic continental shelf water masses,
  J. Phys. Oceanogr. 43(5), 920--940 (2013)

\bibitem{Tsamados2018} Heorton, H.D.B.S., Feltham, D.L. and Tsamados, M.:
  Stress and deformation characteristics of sea ice in a high-resolution, anisotropic sea ice model,
  Philos. Trans. A Math. Phys. Eng. Sci. 376(2129), 20170349 (2018) 
  
\bibitem{FettererUntersteiner} Fetterer, F. and Untersteiner, N.:
  Observations of melt ponds on Arctic sea ice,
  J. Geophys. Res. 103(C11), 24, 821--24, 835 (1998)

\bibitem{HunkeCryo} Hunke, E.C., Notz, D., Turner, A.K. and Vancoppenolle, M.:
  The multiphase physics of sea ice: a review for model developers,
  The Cryosphere 5, 989--1009 (2011)
  
\bibitem{Massonnet} Massonnet, F., Vancoppenolle, M., Goosse, H., Docquier, D., Fichfet, T. and
  Blanchard-Wrigglesworth, E.: Arctic sea-ice change tied to its mean state through thermodynamic processes,
  Nat. Clim. Change 8, 599--603 (2018)
  
\bibitem{Perovich2002aerial} Perovich, D.K., Tucker III, W.B. and Ligett, K.A.:
  Aerial observations of the evolution of ice surface conditions during summer,
  J. Geophys. Res. 107(C10), 8048 (2002)

\bibitem{Hanesiak} Hanesiak, J.M., Barber, D.G., De Abreu, R.A. and Yackel, J.J.:
  Local and regional albedo observations of arctic first-year sea ice during melt ponding,
  J. Geophys. Res. 106(C1), 1005--1016 (2001)
  
\bibitem{Perovich2002albedo} Perovich, D.K., Grenfell, T.C., Light, B. and Hobbs, P.V.:
  Seasonal evolution of the albedo of multiyear Arctic sea ice,
  J. Geophys. Res. 107(C10), 8044 (2002)

\bibitem{Flocco2012} Flocco, D., Schroeder, D., Feltham, D.L. and Hunke, E.C.:
  Impact of melt ponds on Arctic sea ice simulations from 1990 to 2007,
  J. Geophys. Res. 117, C09032 (2012)

\bibitem{Schroeder2014} Schr\"oder, D., Feltham, D.L., Flocco, D. and Tsamados, M.:
  September Arctic sea-ice minimum predicted by spring melt-pond fraction,
  Nat. Clim. Change 4(5), 353--357 (2014)

\bibitem{TaylorFeltham} Taylor, P.D. and Feltham, D.L.: A model of melt pond evolution on sea ice,
  J. Geophys. Res. 109, C12007 (2004)

\bibitem{SkyllingstadPaulson} Skyllingstad, E.D. and Paulson, C.A.: A numerical simulations of melt ponds,
J. Geophys. Res. 112, C08015 (2007) 

\bibitem{Enrico} Rabbanipour Esfahani, B., Hirata, S.C., Berti, S. and Calzavarini, E.: Basal melting driven by turbulent thermal convection,
  Phys. Rev. Fluids 3, 053501 (2018)
  
\bibitem{Luethje} L\"uthje, M., Feltham, D.L., Taylor, P.D. and Worster, M.G.: Modeling the summertime evolution of sea--ice melt ponds,
  J. Geophys. Res. 111, C02001 (2006)

\bibitem{Luethje2} L\"uthje, M., Pedersen, L.T., Reeh, N. and Greuell, W.: Modelling the evolution of supraglacial lakes on the West Greenland ice-sheet margin,
    J. Glaciol. \textbf{52}(179), 608--618 (2006)

\bibitem{SkyllingstadEtAl} Skyllingstad, E.D., Paulson, C.A. and Perovich, D.K.: Simulation of melt pond evolution on level ice, J. Geophys. Res. 114, C12019 (2009)
   
\bibitem{ScottFeltham} Scott, F. and Feltham, D.L.: A model of the three--dimensional evolution of Arctic melt ponds on first--year and multiyear sea ice, 
  J. Geophys. Res. 115, C12064 (2010)
  
\bibitem{Thorndike} Thorndike, A.S., Rothrock, D.A., Maykut, G.A. and Colony, R.: The thickness distribution of sea ice, J. Geophys. Res. 80 (33), 4501--4513 (1975)

\bibitem{CICE} Hunke, E.C., and W.H. Lipscomb: {\it CICE: The Los Alamos Sea Ice Model. Documentation and software user’s manual version 4.0 Tech. Rep. LA-CC-06-012},
  T--3 Fluid Dyn. Group, Los Alamos Natl. Lab., Los Alamos, N. M. (2008)

\bibitem{LIM} Vancoppenolle, M., Bouillon, S., Fichefet, T., Goosse, H., Lecomte, O.,  Morales Maqueda,
  M.A. and Madec, G.:
  {\it The Louvain-la-Neuve sea ice model}, Notes du pole de mod\'elisation, Institut Pierre-Simon Laplace (IPSL),
  Paris, France (2012)
  
\bibitem{FloccoFeltham} Flocco, D. and Feltham, D.L.: A continuum model of melt pond evolution on Arctic sea ice,
  J. Geophys. Res. 112, C08016 (2007)

\bibitem{Flocco2010} Flocco, D., Feltham, D.L. and Turner, A.K.: Incorporation of a physically based melt pond scheme into the sea ice component of a climate model,
J. Geophys, Res. 114, C08012 (2010)

\bibitem{Ahlers} Ahlers, G., Grossmann, S. and Lohse, D.: Heat transfer and large scale dynamics in turbulent Rayleigh-B\'enard convection,
  Rev. Mod. Phys. 81, 503 (2009)

\bibitem{Kim} Kim, J.-H., Moon, W., Wells, A.J., Wilkinson, J.P., Langton, T., Hwang, B., Granskog, M.A. and Rees Jones, D.W.:
  Salinity control of thermal evolution of late summer melt ponds on Arctic sea ice,
  Geophys. Res. Lett. 45, https://doi.org/10.1029/2018GL078077 (2018)
  
\bibitem{GL} Grossmann, S. and Lohse, D.: Scaling in thermal convection: a unifying theory,
  J. Fluid Mech. 407, 27--56 (2000)

\bibitem{Malkus} Malkus, M.V.R.: The heat transport and spectrum of thermal turbulence,
  Proc. R. Soc. London A 225, 196 (1954)

\bibitem{Domaradzki} Domaradzki, J.A. and Metcalfe, R.W.: irect numerical simulations of the effects of shear on
  turbulent Rayleigh-B\'enard convection,
  J. Fluid Mech. 193, 499 (1988)

\bibitem{Rallabandi} Rallabandi, B., Zheng, Z., Winton, M. and Stone, H.A.: Formation of sea ice bridges in narrow
  straits in response to wind and water stresses,
  J. Geophys. Res. Oceans 122(7), 5588--5610 (2017)
  
\bibitem{Scagliarini} Scagliarini, A., Gylfason A. and Toschi, F.: Heat-flux scaling in turbulent Rayleigh-B\'enard convection
  with an imposed longitudinal wind,
  Phys. Rev. E 89, 043012 (2014)

\bibitem{Prandtl} Prandtl, L.: Bericht \"uber die Entstehung der Turbulenz,
  Z. Angew. Math. Mech, 5, 136--139 (1925)
  
\bibitem{Tsamados2015} Tsamados, M., Feltham, D.L., Petty, A.A., Schr\"oder D. and Flocco, D.:
  Processes controlling surface, bottom and lateral melt of Arctic sea ice in a state of the art sea ice model,
  Phil. Trans. R. Soc. A 373, 20140167 (2015)
  
\bibitem{Eicken2002} Eicken, H., Krouse, H.R., Kadko, D. and Perovich, D.K.: Tracer studies of pathways and rates
  of meltwater transport through Arctic summer sea ice,
  J. Geophys. Res. 107(C10), 8046 (2002)

\bibitem{Landau} Landau, L.D. and Lifshitz, E.M.: Fluid Mechanics (Second Edition), Pergamon Press (1987)

\bibitem{Oron} Oron, A., Davis, S.H. and Bankoff, S.G.: Long-scale evolution of thin liquid films,
  Rev. Mod. Phys. 69(3), 931--980 (1997)

\bibitem{Marche} Marche, F.: Derivation of a new two-dimensional viscous shallow water model with varying topography,
  bottom friction and capillary effects,
  Eur. J. Mech. B Fluids 26, 49--63 (2007)
  
\bibitem{Hvidegaard} Hvidegaard, S.M. and Forsberg, R.: Sea-ice thickness from airborne
  laser altimetry over the Arctic Ocean north of Greenland,
  Geophys. Res. Lett., 29(20), 1952 (2002)

\bibitem{HK} Hoshen, J. and Kopelman, R.: Percolation and cluster distribution. I. Cluster multiple labeling technique
  and critical concentration algorithm, Phys. Rev. B. 14, 3438--3445 (1976)

\bibitem{MoritzEtAl} Moritz, R.E., Curry, J.A., Thorndike, A.S. and Untersteiner, N.:
  {\it SHEBA a Research Program on the Surface Heat Budget of the Arctic Ocean},
  Rep. 3, 34 pp., Arctic Syst. Sci.: Ocean-Atmos.-Ice Interact. (1993)

\bibitem{MoritzPerovich} Moritz, R.E. and Perovich, D.K., (Eds.), {\it Surface Heat Budget of the Arctic Ocean, Science Plan, ARCSS/OAII},
  Rep. 5, 64 pp., Univ. of Wash., Seattle, Wash. (1996)

\bibitem{Ising} Ma, Y.-P., Sudakov, I., Strong, C. and Golden, K.M.: Ising model for melt ponds on Arctic sea ice, arXiv:1408.2487 (2014)

\bibitem{Popovic} Popovi\'c, P., Cael, B.B., Silber, M. and Abbot, D.S.: Simple rules govern the patterns of arctic sea ice melt ponds, Phys. Rev. Lett. 120, 148701 (2018)

\bibitem{PerovichEtAl2009} Perovich, D.K., Grenfell, T.C., Light, B., Elder, B.C., Harbeck, J., Polashenski, C., Tucker III, W.B.,
  and Stelmach, C.:
  Transpolar observations of the morphological properties of Arctic sea ice, J. Geophys. Res., 114, C00A04 (2009)
  
\bibitem{Hohenegger} Hohengger, C., Alali, B., Steffen, K.R., Perovich, D.K. and Golden, K.M.: Transition in the fractal geometry of Arctic melt ponds, The Cryosphere, 6,
  1157--1162 (2012)
  
\bibitem{Mandelbrot} Mandelbrot, B.B.: {\it The fractal geometry of Nature}, Freeman, New York (1982)

\bibitem{ChengMatGeo} Cheng, Q.: The perimeter-area fractal model and its application to geology,
  Math. Geol. 27, 69--84 (1995) 

\bibitem{Grassberger} Grassberger, P.: Generalized dimensions of strange attractors,
  Phys. Lett. 97A(6), 227--230 (1993)
  
\end{thebibliography}
%

\end{document}